\documentclass[prb, showpacs, twocolumn,superscriptaddress]{revtex4-1}

\usepackage{amsmath}
\usepackage{amssymb}
\usepackage{graphicx}
\newcommand{\be}{\begin{equation}}
\newcommand{\ee}{\end{equation}}

\begin{document}

\title{The extraordinary importance of self-avoiding behavior in two-dimensional polymers: Insights from large-deviation theory}

\author{Eleftherios Mainas}
\affiliation{Department of Chemistry, Brown University, Providence, RI 02912}
\affiliation{Department of Chemistry, University of North Carolina, Chapel Hill, North Carolina 27599}

\author{Jan Tobochnik}
\affiliation{Department of Physics, Brown University, Providence, Rhode Island 02912}
\affiliation{Department of Physics, Kalamazoo College, Kalamazoo, Michigan 49006}

\author{Richard M. Stratt}
\email{richard\_stratt@brown.edu}
\affiliation{Department of Chemistry, Brown University, Providence, RI 02912}

\date{\today}
\keywords{}

\begin{abstract}
Some recent work pointed out the usefulness of taking a large-deviation perspective when trying to extract anything resembling a macroscopic order parameter from a computer simulation.  In this paper we note that the end-to-end distance of polymers is such an order parameter.  The presence of long-ranged excluded volume interactions leads to significant qualitative differences between the conformations of two- and three-dimensional polymers, some of which are difficult to quantify in computer simulations of realistic (off-lattice) polymer models.  But we show here that phenomena such as the greatly enlarged non-Hooke’s-law elasticity present in 2D are straightforward to extract from simulation using a large-deviation framework – even though simulating that nonlinearity is tantamount to simulating a 4th order susceptibility.  The large-deviation perspective includes both a set of thermodynamic-like tools suitable for studying finite-size systems and a realization that an accurate description of the system’s average behavior needs to be consistent with how improbably large fluctuations would behave in that system.  The latter is key because strong correlations are absent in this asymptotic large fluctuation regime, so the regime’s far-reaching effects can be analytically incorporated into the analysis of simulation data.  That, in turn, allows us to direct the efforts of simulations away from difficult-to-sample rare-event domains.  We illustrate this point with two- and three-dimensional Monte Carlo simulations (and exact results) on two models of a single isolated polymer chain: a chain of linked hard spheres, which has long-ranged excluded volume effects, and a discretized worm-like chain, which does not.
\end{abstract}

\pacs{}

\maketitle

\section{Introduction} 
This paper re-examines the old problem of the conformational statistics of single, isolated, polymer chains,~\cite{A,a,b} and in particular, the unusual conformational statistics occurring in two dimensions.\cite{c,c1,d}  Any realistic polymer chain model has a certain self-avoiding random-walk character,~\cite{A,e} but the long-ranged excluded volume effects responsible for that self-avoidance turn out to have a much more fundamental effect on conformation in two dimensions than they do in higher dimensions.~\cite{e1} The topological constraints imposed by two dimensions result in polymers exploring conformational states with significantly larger sizes than expected for an ordinary (non-self-avoiding) random walk.  The resulting large size fluctuations lead, in turn to two-dimensional conformational distributions that are fundamentally non-Gaussian, even in the infinite-chain limit.

None of this phenomenology is news {\it per se}.  Field theoretical treatments\cite{f,f1} and computer simulations of self-avoiding random walks on lattices~\cite{c1} have provided numerous convincing demonstrations.  In particular, the analogy between self-avoiding walks on a lattice and the unintuitive, but mathematically well-defined, zero-component limit of n-component spins on a lattice~\cite{A} allows one to show that, in $D$ dimensions, the Flory length $R_F$ (the root-mean-square of the end-to-end distance vector ${\bf R}$) scales with $N$, the number of random-walk steps, in a fashion very close to that predicted by Flory’s own arguments from more than half a century ago~\cite{g,h,i}
\begin{equation}
R_F = \langle R^2 \rangle^{1/2} \sim N^{\nu}, \,\,\,\,\,\,   \nu({\rm Flory}) = \frac{3}{D+2},   \label{eq1.1} 
\end{equation}
The conventional random walk ($N^{1/2}$) scaling expected from a Gaussian distribution~\cite{j} is not reached until $D = 4$, but as has been noted in the literature~\cite{k} (and we discuss later) the special ($N^{3/4}$)) scaling found in two dimensions leads to noticeably distinct physical behavior. What we want to explore in this paper is a somewhat different way of understanding the implications of this scaling in physical, off-lattice, models of polymers.

To help motivate our approach, it is worth placing this problem in a more general context.  Understanding the conformational tendencies of simple polymers is, in many ways, a special case of the larger set of problems posed by {\it orientational ordering}:  The states of the $N$ fundamental units (polymer links, individual molecules, or lattice spins) are described by individual unit vectors ${\hat \Omega}_j \,\, (j = 1,2,...N)$ and the physical questions revolve around the likely values of an orientational order parameter.  The order parameters differ in detail, but they generally take the form of a sum over the orientations of all the component units measured relative to some unit vector ${\hat h}$,
\begin{equation}
M = \sum_{j=1}^N F({\hat \Omega}_j \cdot {\hat h}). \label{eq1.2} 
\end{equation}
where $F(u)$ is some physically relevant function with $F(1) = 1$.  For simple polymer chains with fixed bond lengths $d$, we can take the  ${\hat \Omega}_j$ to be monomer unit vectors, making the order parameter (with $F(u) = u$) proportional to the component of the end-to-end distance vector
\begin{equation}
{\bf R} = d\sum_{j=1}^N {\hat \Omega}_j \label{eq1.3} 
\end{equation}
along some pre-determined direction  ${\hat h}$.  Similarly, (keeping $F(u) = u$), if we take the unit vectors to be classical spins and take  ${\hat h}$ to be the direction of an applied magnetic field then $M$ in Eq.~(\ref{eq1.2}) is the longitudinal magnetization of a classical magnetic system.

More general functions $F(u)$ allow for yet other possibilities. For liquid crystals formed from linear molecules in $D$ dimensions, with  the ${\hat \Omega}_j$ the orientational unit vectors of the individual molecules and  ${\hat h}$ the director of that order, choosing
\begin{equation}
F(u) = \frac{Du^2 -1}{D-1}  \label{eq1.4} 
\end{equation}
makes the order parameter the standard measure of the extent of nematic ordering.~\cite{l}
	
In each case, what distinguished orientational ordering from more general situations is that orientational order {\it saturates}.  The largest possible value of  
$m = M/N$ the order per degree of freedom, is always unity: polymers with fixed bond lengths cannot be stretched beyond the point that all the bonds are parallel, magnets cannot be magnetized beyond the point of perfect ordering, and so forth.  In the course of some previous work exploring liquid crystals,~\cite{l} we were able to show that combining the analytical implications of this saturation feature with a large-deviation-theory perspective~\cite{m} allowed us to skirt some of the rare-event sampling issues common to computer simulation-derived measurements of ordering.  The goal in this paper is to take these same perspectives and apply them to computer simulations of polymer conformations.
	
	Our previous efforts~\cite{l} were motivated by a desire to understand the occurrence of improbably large regions of local nematic ordering in putatively disordered liquid crystalline systems – a classic rare-event sampling problem.  What we observed was, first, that even though a situation with perfect orientational order could never be reached in any practical simulation of liquid crystals, the fact that such saturation behavior is mathematically possible meant that there had to be a pole in the order-parameter-versus-applied-ordering-field equation of state at $m = 1$; the applied field had to diverge to achieve perfect ordering.  That, in turn, means the free energy had to have a logarithmic singularity at $m = 1$, a situation with significant ramifications for the behavior of the free energy throughout the entire range of possible orderings.  Secondly, we observed that the statistical mechanical details of the saturation end of the range were remarkably simple.  For large enough applied fields, the field dominates the intermolecular interactions, making the orientational behavior largely that of isolated molecules in the presence of a field.~\cite{l}
	
	The other end of the range of ordering possibilities, that corresponding to weak applied fields and strong intermolecular correlations, poses much more of a statistical mechanical challenge.  But that regime is also the one that corresponds to small fluctuations, making it straightforward to investigate via simulation.  Since large deviation theory suggests how one can accurately extrapolate between these regimes, we were able to devise a way to take computer simulation information from situations in which sampling was reliable and use it to help understand improbably large and difficult-to-sample ordering fluctuations.~\cite{l}
	
	The basic idea of the present paper is to apply much the same logic to finding the end-to-end distance distribution of polymer chains in any number of dimensions.  Viewed from the lens of large-deviation theory, this distribution is equivalent to predicting the polymer force-extension equation of state: the equilibrium relationship between the end-to-end distance and the applied tensile force required to stretch the polymer to achieve that distance.~\cite{n} Understanding such equations of state has become increasingly relevant because of the ability to experimentally measure such relationships for DNA molecules.~\cite{o,p,q,r}  However, for our present purposes, the virtues of working with equations of state are that they are straightforward to find using computer simulation and, more importantly, that they provide a particularly transparent way to quantify the intrinsically non-Gaussian character that shows up in physical two-dimensional polymers.  We shall show, in particular, how our large-deviation formalism explains why nonlinear elasticity is present in two-dimensional polymers with realistic long-ranged excluded volume effects while it is absent in both the three-dimensional analogue and in two-dimensional polymers with just local excluded volume.

As in our previous work,~\cite{l} the formalism is not intended to be a first-principles analytical prediction using nothing but the molecular Hamiltonian, but rather to offer guidance on how to extract interesting but difficult to compute, information from computer simulation.  Unlike our liquid-crystal studies, the principal issue now is not infrequent events (which, in the polymer context would involve unusually large polymer extensions associated with large amounts of applied stress) but the nonlinear polymer response seen under weak stress conditions.  The reason that measuring this response poses a computational challenge is that it necessitates simulating a 4th order susceptibility, an intrinsically noisy target.  Our analysis shows how the information in such quantities can be extracted from simulations without computing them directly.
	
	To this end, the remainder of this paper will be organized as follows: Section~\ref{sII} will describe how large-deviation theory converts the statistics of finite-sized polymer end-to-end distances into a problem of knowing the connection between those distances and an applied force.  It then shows how deducing that connection from simulation is aided by what we know analytically about the large-extension limit.  Section~\ref{sIII} demonstrates that combining these results with the known scaling of the mean-square end-to-end distance is sufficient to help us understand the special character of polymers confined to two dimensions.  Section~\ref{sIV} then presents the model polymer systems that we test our predictions on, discussing, in particular, the similarities and differences between the discretized worm-like chain model and physical polymers with long-ranged excluded-volume effects.  Section~\ref{sV} tests our formalism and its predictions by simulation.  We conclude in Sec.~\ref{sVI}.

\section{LARGE-DEVIATION THEORY AND POLYMER CONFORMATIONAL FLUCTUATIONS}\label{sII}

Suppose we limit ourselves to thinking about simple polymer chains of the sort described in the introduction, those with $N$ monomers each of fixed length $d$, and ask what the probability distribution is of the end-to-end vector ${\bf R}$ given by Eq.~(\ref{eq1.3}).  Note that since we will be assessing how Gaussian this distribution is, it will be important for us not to build in any assumptions at this point that imbue the monomer lengths themselves with Gaussian statistics.~\cite{s}  We will further assume that, in addition to whatever other intra-chain interactions there are, there will be excluded volume interactions preventing the chain from crossing through itself at any point.

Much as we described in our previous work on liquid crystals,~\cite{l} the G\"artner-Ellis large-deviation theorem~\cite{m}  tells us that, regardless of the interaction details, the Helmholtz free energy per particle governing the probability density of the extension ${\bf R}$ (commonly called the ``rate function"  $I({\bf R})$ in the jargon of the field) can be derived from what is essentially the Gibbs free energy per particle $\lambda({\bf f})$, where ${\bf f}$ is the applied tensile force (over  $k_B T$ ) conjugate to ${\bf R}$:
\be
e^{N \lambda({\bf f})}= \langle e^{{\bf f} \cdot {\bf R}} \rangle, \label{eq2.1}
\ee
\be
N I({\bf R}) = {\bf f}^* \cdot {\bf R} - N \lambda({\bf f}^*) \label{eq2.2}
\ee
with ${\bf f}^*$ the particular force vector satisfying the relation
\be
{\bf R} =  [\nabla_{\bf f}  \lambda({\bf f}) ]_{{\bf f} = {\bf f}^*} \,\, .   \label{eq2.3}
\ee
(Brackets of the form shown in Eq.~(\ref{eq2.1}) are meant to indicate canonical averages governed by the Hamiltonian of the polymer in the absence of any external forces.)
	
	This mathematical structure is, of course, precisely what one finds from thermodynamics, complete with the quintessentially thermodynamic pattern of having a natural association of a putatively extensive variable ${\bf R}$ with a conjugate (putatively) intensive variable $f$.  Indeed, if we take the force to be applied in a fixed direction   and ask for the free energy governing the parallel component of the (scaled) end-to-end length, $x$ (with the subscripted brackets indicating averages in the presence of the applied force),
\begin{eqnarray}
{\bf f} &=& (f/d) {\hat f},   \nonumber \\
x &=& \frac{1}{Nd} \langle {\bf R} \cdot {\hat f} \rangle_f  = \frac{1}{N} \biggl < \sum_{j = 1}^N {\hat \Omega}_j \cdot {\hat f} \biggr>_f \,\,  , (0 \le x \le 1) \label{eq2.4}
\end{eqnarray}
combining Eqs.~(\ref{eq2.2}) and (\ref{eq2.3}) and integrating by parts shows that the free energy  $I(x)$  is simply the reversible work required to achieve an extension $x$,
\be
I(x) = \int_0^x f(x')dx'  ,  \label{eq2.5}
\ee
a quantity that can be evaluated immediately once we know the {\it equation of state} $f = f(x)$.~\cite{t}

So why not simply rely on thermodynamics to do our calculations?  The reason is that large deviation theory allows us to explore the scaling of our system with $N$, rather than assuming the traditional, thermodynamic, large-$N$ limit. As long as we know the equation of state for a given value of $N$, we can write the free energy for that $N$ value.  But whether truly thermodynamic or not, we still need to know the equation of state to make use of Eq.~(\ref{eq2.5}).  Here is where a second observation made in our previous work enters.
For large enough applied forces, we know that intra-polymer forces become negligible, meaning that the equation of state must tend to the ideal polymer equation of state, that of a freely-jointed chain.  In that limit, the free energy in Eq.~(\ref{eq2.1}) is thus just a sum of independent contributions from each polymer link orientation
 ${\hat \Omega}_j$,
\be
\lambda(f) = \frac{1}{N} \ln {\bigl< e^{ f \sum_{j=1}^N {\hat f} \cdot {\hat \Omega}_j} \bigr>} = \ln {\biggl( \int d{\hat \Omega} e^{f({\hat f} \cdot {\hat \Omega})} } \biggr) , 
\ee
making the corresponding ``inverse" equation of state, Eq.~(\ref{eq2.3}), in any number of dimensions $D$,
\begin{eqnarray} 
x(f) &=&  \frac{\partial \lambda(f)}{\partial f} = \frac{\int_0^{\pi} \, d \theta  \sin^{D-2} {\theta} e^{f \cos{\theta}} }{ \int_0^{\pi} \, d \theta  \sin^{D-2} {\theta} }  \nonumber \\
&=& \Gamma \bigg(\frac{D}{2}\bigg) \frac{I_{\frac{D}{2} -1} (f)}{ I_{\frac{D}{2}} (f)}.   \label{eq2.6}
\end{eqnarray}
with $I_n(f)$ denoting the $n$-th order modified Bessel functions.  For example, in one, two, and, three dimensions, we have the familiar results that
\be
x = \begin{cases} \tanh{(f)} &(1d) \\  I_1(f)/I_0(f) &(2d) \\  \coth{(f)} - \frac{1}{f} &(3d).\end{cases}\label{eq2.7}
\ee

The known asymptotic behavior of Bessel functions~\cite{u} therefore tells us what we should expect for the behavior of our polymer chain as it approaches its maximum length $x=1$.  In any dimension $D$, we must have the limiting behavior that as $f \to \infty$,~\cite{v}
\be
x(f) = 1 - \frac{D-1}{2f} + O\biggl(\frac{1}{f^2}\biggr).
\ee
or, equivalently as $x \to 1$, the applied force must obey the requirement that,
\be
f(x)  =  \frac{D-1}{2} \frac{1}{1-x} + O(1-x)^0.  \label{eq2.8} 
\ee

The opposite limit, that of {\it minuscule applied forces and small extensions}, will reflect the full impact of whatever intra-chain correlations are present. But in the limit that the extensions (the chain fluctuations) are small, the free energy Eq.~(\ref{eq2.1}) is governed by a low-order cumulant expansion
\be
N\lambda(f) = \langle {\bf R} \cdot {\bf f} \rangle  + \frac{1}{2} \bigl<(\delta {\bf R} \cdot {\bf f})^2 \bigr> + \ldots
\ee
with $\delta {\bf R} = {\bf R} - \langle{\bf R} \rangle$   the extension fluctuation.  Because symmetry dictates that, in the absence of an applied force, the average extension vector $\langle{\bf R} \rangle = 0$,
\begin{eqnarray}
\lambda(f) &=& \frac{1}{2N} f^2 \bigr< (({\bf R}/d) \cdot {\hat f} )^2 \bigr> + \ldots \nonumber \\
&=&  \frac{1}{2ND} f^2 \bigr<  ({\bf R}/d)^2  \bigr> + \ldots \nonumber \\
&=& \frac{f^2}{2D} \chi + \ldots, \label{eq2.9} 
\end{eqnarray}
where we have defined the {\it susceptibility}  $\chi$ to be
\be
\chi = \frac{1}{N} \biggl< \sum_{j,k=1}^N {\hat \Omega}_j \cdot {\hat \Omega}_k \biggr>. \label{eq2.10} 
\ee
This (central-limit-theorem) behavior therefore implies that the corresponding inverse equation of state in the $f \to 0$ limit must be
\be
x(f) = \frac{\partial \lambda(f)}{\partial f} = \biggl(\frac{\chi}{D} \biggr)f  + O(f^2) \label{eq2.11} 
\ee
so that the applied force in the small extension limit must obey Hooke’s law
\be
f(x) = \biggl(\frac{D}{\chi} \biggr) x + O(x^2). \label{eq2.12} 
\ee

We are not all that concerned with either of the extreme limits shown in Eqs.~(\ref{eq2.8}) and (\ref{eq2.12}), but we know that the exact, full, equation of state is compelled to interpolate between them.  That is, we have the rigorous requirement that in $D$ dimensions
\be
f(x) = \begin{cases} \bigr(\frac{D}{\chi}\bigr)x &x \to 0 \\ \frac{D-1}{2} \frac{1}{1-x} + \eta &x \to 1 \end{cases} \label{eq2.13}
\ee
with $\eta$ a constant.  As we observed in our earlier work,~\cite{l} the very simplest interpolated equation of state consistent with both these limiting behaviors and symmetry requirements (in our case, that the end-to-end extension vector ${\bf R} \to -{\bf R}$ whenever the applied force vector ${\bf f} \to -{\bf f}$) is the Pad\'e form featuring a pole at the full extension $x \to 1$ limit~\cite{w}
\begin{eqnarray}
f(x) &=& \biggl(\frac{D}{\chi}\biggr) \frac{x + Ax^3 + Bx^5}{1-x^2} \label{eq2.14} \\
A &=& -2 + \biggl(\frac{\chi}{D}\biggr) \bigl (\eta + \frac{9}{4} (D-1) \bigr) \nonumber\\
B &=& 1-  \biggl(\frac{\chi}{D}\biggr) \bigl (\eta + \frac{5}{4} (D-1) \bigr). \nonumber
\end{eqnarray}
Large deviation theory, Eq.~(\ref{eq2.5}), then tells us the simplest possible free energy consistent with all of our conditions must have the form
\begin{eqnarray}
 I(x) &=& \frac{1}{2} \Delta x^2 - \frac{1}{4} \Gamma x^4 - \frac{(D-1)}{2} \ln {(1-x^2)} \label{eq2.15} \\
\Delta &=&  \biggl(\frac{D}{\chi}\biggr)  - (D-1)  , \,\,\,\,\,  \Gamma = \Delta -  \eta  - \biggl(\frac{D-1}{4}\biggr) \nonumber
\end{eqnarray}
complete with the corresponding logarithmic singularity at full extension that we alluded to in the Introduction.

We will put Eq.~(\ref{eq2.14}) (and by extension, Eq.~(\ref{eq2.15})) to a numerical test in Sec.~\ref{sV}, treating $\chi$ and $\eta$  as quantities  determined by simulation.  However, probably the most interesting feature of these equations is that they suggest some immediate scaling predictions. 

\section{SOME SCALING IMPLICATIONS}\label{sIII}

\subsection{Relative scaling of the terms in the free energy}

Although carrying out our analysis necessitated thinking through what polymers do near their full extension ($x = 1$) limit, our ultimate goal is to understand more typical conformational states, those whose lengths $R = |{\bf R}|$   are close to the Flory length  $R_F$  specified by Eq.~(\ref{eq1.1}).~\cite{A}  Because Eq.~(\ref{eq1.3}) and (\ref{eq2.10}) tells us that the susceptibility $\chi$ itself is actually proportional to the mean-square end-to-end displacement
\be
R_F^2 = \langle{\bf R}^2 \rangle = Nd^2 \chi  \implies x \sim (R/R_F) \sqrt{\chi/N}\,\,\,. \label {eq3.1} 
\ee
That is, given the known scaling with $N$ implied by Eq.~(\ref{eq1.1}),  $\chi \sim N^{2\nu-1}$, we find  $x\sim N^{\nu-1}$.  In other words, for high polymers in any dimension $D > 1$, the most probable situations will be those in which $x \ll 1$. (In 1d $\nu=1$; the presence of excluded volume means that $x$ is always identically equal to 1.)

Under these circumstances, the free energy per monomer predicted by large deviation theory, Eq.~(\ref{eq2.15}), can be written as
\begin{eqnarray}
I(x) &=& \frac{1}{2} \bigl(\Delta+ (D-1)\bigr)x^2  - \frac{1}{4} \bigl(\Gamma - (D-1)\bigr) x^4 + \ldots. \nonumber \\
&=& \frac{1}{2} \biggl(\frac{D}{N}\biggr) \biggl(\frac{R}{R_F}\biggr)^2 \nonumber \\
&-& \frac{1}{4} \bigl(\Gamma-(D-1)\bigr) \biggl(\frac{\chi}{N}\biggr)^2 \biggl(\frac{R}{R_F}\biggr)^4 +\ldots \nonumber \\
&=& O(N^{-1})\biggl(\frac{R}{R_F}\biggr)^2  \nonumber \\&-& O(N^{4(\nu-1)})\biggl(\frac{R}{R_F}\biggr)^4 + \ldots   \label{eq3.2} 
\end{eqnarray}
because  we expect $\Gamma = O(N^0)$.~\cite{x}

Now let us consider the implications of this formula when $R/R_F$ is on the order of 1.  In four or higher dimensions, polymers are ideal, at least in the sense that $\nu = \frac{1}{2}$, so we would predict that the $O(N^{-1})$ Gaussian $(R/R_F)^2$ term will always dominate the $O(N^{-2})$ quartic term.  Similarly, in three dimensions, the (nearly exact) Flory prediction is that $\nu = \frac{3}{5}$, so we would expect that the now  $O(N^{-8/5})$ quartic term would still not be sufficient to prevent Gaussian behavior for large enough polymers.  However, in two dimensions, the Flory $\nu = \frac{3}{4}$ (which is actually exact)~\cite{h} turns out to be precisely what is needed to make both the Gaussian and quartic terms $O(N^{-1})$.  We would therefore expect that the conformational statistics of two-dimensional polymers to be unique in that, unlike their higher-dimensional analogues, the 2D case will always be fundamentally non-Gaussian, even in the large $N$ limit.
	
	It is worth examining the origins of this prediction in a little more detail.  Our large-deviation extrapolation scheme is effectively proposing that all of the scaling behavior in the polymer conformation problem is encapsulated in the  $O(N^{2\nu-1})$ scaling of the susceptibility $\chi$. But, this apparent limitation of the formalism actually builds in the exact behavior
of this particular problem; the large-$N$ limit of a self-avoiding walk acts as a
critical point controlled by the single critical exponent, $\nu$.~\cite{x1}  For ideal systems,  $\chi \sim N^0$ (and is, in fact, identically 1 for a freely-jointed polymer, one with no intra-chain interactions), but adding in long-ranged self-avoiding interactions in any dimension less than four generates an exponent $\nu > \frac{1}{2}$, making $\chi$ grow with increasing chain size.  Still, it is apparently not until one is confined to two dimensions that the non-Gaussian character becomes really manifest.
	
\subsection{Just how non-Gaussian are our polymers?} 

We can quantify this loss of Gaussian behavior by defining a non-Gaussian parameter, $\alpha_{NG}$ in the same fashion that is done in many other condensed matter contexts.~\cite{y}  If our D-dimensional end-to-end vector ${\bf R}$ did obey Gaussian statistics, it would have a probability density
\be
P({\bf R}) \sim e^{-\frac{D}{2}R^2/\langle R^2 \rangle},     \,\,\,\,\,\,\,\,\,\, \int d{\bf R} P({\bf R}) = 1\,\,\, .
\ee
Because it is straightforward to show that for such a probability density
\be
\frac{\langle R^4 \rangle}{\langle R^2 \rangle^2} = \frac{D+2}{D},
\ee
we will define the dimensionless non-Gaussian parameter for our actual polymer system to be
\be
\alpha_{NG} = \frac{\langle R^4 \rangle}{\langle R^2 \rangle^2} - \frac{D+2}{D}.  \label{eq3.3} 
\ee

There are, of course, other ratios of moments that one could also use as non-Gaussian metrics, but one can show that what this particular parameter measures physically is the leading order non-linear elasticity of a polymer.  Because of the afore-mentioned inversion symmetry of polymer end-to-end vectors, the equation of state must have its leading deviations from (linear) elastic behavior, Eq.~(\ref{eq2.11}), be of the form
\be
x(f) = \biggl(\frac{\chi}{D}\biggr) f + af^3 + O(f^5) ,  \label{eq3.4} 
\ee
and one can show that our $\alpha_{NG}$ is directly proportional to the inelasticity constant $a$.

To see this result, notice that for any isotropically distributed vector ${\bf R}$  in $D$ dimensions
\be
\langle R^2 \rangle = \frac{\langle ({\bf R} \cdot {\hat f})^2 \rangle}{\langle \cos^2{\theta} \rangle} = D \langle ({\bf R} \cdot {\hat f})^2 \rangle, \label{eq3.5} 
\ee
\be
\langle R^4 \rangle = \frac{\langle ({\bf R} \cdot {\hat f})^4 \rangle}{\langle \cos^4{\theta} \rangle} = \frac{D(D+2)}{3} \langle ({\bf R} \cdot {\hat f})^4\rangle, \label{eq3.6} 
\ee
with $\theta$ equal to the angle between ${\bf R}$  and any arbitrarily chosen unit vector ${\hat f}$.  We can therefore rewrite Eq.~(\ref{eq3.3}) in terms of fluctuations of the (scaled) parallel component of the polymer’s end-to-end length, $M$
\be
\alpha_{NG} = \frac{D+2}{3D} \biggl( \frac{\langle M^4 \rangle - 3 \langle M^2 \rangle^2}{ \langle M^2 \rangle^2} \biggr), \label{eq3.7}
\ee
where
\be
{\bf M} = \frac{{\bf R} \cdot {\hat f}} {d} = {\hat f} \cdot \sum_{j=1}^N {\hat \Omega}_j . \label{eq3.8}
\ee
Inasmuch as the Hamiltonian of our polymer in the presence of an applied force $f$ is given by
\begin{eqnarray}
-\beta {\cal H} &=& -\beta {\cal H}_0 +{\bf F} \cdot {\bf R}  \nonumber \\ 
&=& -\beta {\cal H}_0 + f M,   \,\,\,\,\,\,  \beta = 1/(k_BT), \label{eq3.9}  
\end{eqnarray}
with ${\cal H}_0$  the original (force-free) Hamiltonian, it is clear that successive derivatives of the equation of state, Eq~(\ref{eq2.4}),
\be
x(f) = N^{-1} \langle M \rangle_f, \label{eq3.10}
\ee
with respect to $f$ yield successively higher-order susceptibilities measuring parallel-component fluctuations
\begin{eqnarray}
\chi_2 &=& \biggl( \frac{\partial \langle M \rangle_f }{\partial f} \biggr)_{f = 0} = \langle M^2 \rangle, \label{eq3.11}\\
\chi_4 &=& \biggl( \frac{\partial^3\langle M \rangle_f }{\partial f^3} \biggr)_{f = 0} = \langle M^4\rangle - 3\langle M^2\rangle^2, \label{eq3.12}
\end{eqnarray}
Thus, we see that our non-Gaussian parameter, Eq.~(\ref{eq3.7}), is fundamentally a probe of the 4th-order fluctuations
\be
\alpha_{NG} = \biggl(\frac{D+2}{3D}  \biggr) \biggl(\frac{\chi_4}{\chi_2^2}  \biggr).  \label{eq3.13}
\ee
From Eqs.~(\ref{eq2.10}), (\ref{eq3.5}), (\ref{eq3.8}), and (\ref{eq3.11}),
\be
\chi_2 = \frac{N}{D} \chi,
\ee
and from Eqs.~(\ref{eq3.4}), (\ref{eq3.10}), and (\ref{eq3.12}),
\be
\chi_4 = N \biggl( \frac{\partial^3 x}{\partial f^3} \biggr)_{f=0} = 6aN,
\ee
we can also see that those fluctuations are a direct probe of the leading-order non-linear elasticity
\be
\alpha_{NG}  = \frac{2D(D+2)}{N\chi^2} a.   \label{eq3.14}
\ee

The scaling of this $\alpha_{NG}$ is revealing.  For a freely-jointed polymer, both $\chi$ and $a$  are independent of $N$, so it is evident that the non-Gaussian parameter vanishes at large $N$. In particular, after computing $a$ by triple differentiating the freely-jointed equation of state, Eq.~(\ref{eq2.6}), we find that
\be
\alpha_{NG}(\rm freely\mbox{--}jointed\,\,chain)  = -\frac{2}{ND}.   \label{eq3.15}
\ee
Thus, in a very literal sense, freely-jointed chains are fundamentally Gaussian in any dimension.
What happens when a polymer has long-ranged excluded volume interactions? Because we now have our interpolated prediction for the equation of state of such polymers, we can make a prediction.  After inverting Eq. (\ref{eq2.14}) to obtain a power series for $x(f)$ and differentiating, we find
\be
a = -\biggl (\frac{\chi}{D} \biggr)^3(1+A) .   \label{eq3.16}
\ee
Remembering that in the presence of long-ranged excluded volume effects $ \chi \sim N^{2\nu-1}$, Eq. (\ref{eq3.14}) predicts
\be
\alpha_{NG}  \sim - \frac{\chi^2}{N} = O(N^{4\nu-3}).   \label{eq3.17}
\ee
That is, at small extensions, our analysis leads us to expect fundamentally linear elasticity in three dimensions, but not in two:
\be
\alpha_{NG} = \begin{cases} O(N^0) &{\rm in \,\,2d} \\ O(N^{-3/5}) &{\rm in \,\, 3d}   \\ O(N^{-1}) &{\rm in \,\, 4d.} \end{cases} \label{eq3.18}
\ee
We turn next to checking the validity of this prediction, and of our equation of state form more generally, in the context of some specific polymer models.

\section{APPLICATION TO SOME SIMPLE POLYMER MODELS}\label{sIV}

The fact that any physically realizable polymer chains are going to have repulsive intra-chain interactions present at all length scales lies at the heart of the deviation from simple random-walk scaling discussed in the last section.  Perhaps the simplest $N$-link polymer model that features these realistic long-ranged self-avoidance properties and has a maximum possible extension is the {\it hard-sphere chain}.~\cite{z}  This paper will consider chains consisting of $N+1$ hard spheres, each of diameter $\sigma$ with the distance between neighbor spheres fixed at a distance $d$ (fixing the maximum possible chain length to be $Nd$), but no restriction on polymer-link/polymer-link angles other than those imposed by the hard spheres.  We will deal only with the case $d = \sigma$, although any value of  $d \le \sigma$ would prevent such a chain from passing through itself.

We will discuss our simulations of this model and our application of the formalism of the last two sections to this model in Sec.~\ref{sV}.  But it will prove interesting to compare what we find with the results of a model in which the intra-chain interactions are limited to more microscopic lengths.  To do so, we will also look at what we shall call the discretized worm-like chain model.~\cite{aa,bb,cc} What one typically refers to as the worm-like chain picture is a continuum model that builds in the elastic costs of bending a stiff polymer chain (without including the long-ranged interaction that prevents polymers from crossing themselves).~\cite{dd,ee,ff,gg,ii,jj,kk,ll,mm}  However, it is possible to show that when the standard continuum model is discretized, it becomes a chain of $N$ $D$-dimensional unit vectors ${\hat \Omega}_j$  interacting with nearest-neighbor attractions.  In fact, in the presence of an applied tensile force applied in a direction ${\hat f}$ , the Hamiltonian is just
\be
-\beta {\cal H} = K \sum_{j=2}^N {\hat \Omega}_{j-1} \cdot {\hat \Omega}_j + f \sum_{j=1}^N {\hat \Omega}_j \cdot {\hat f}, \label{eq4.1}
\ee
where $K = \ell_p/d$   is the ratio of the polymer’s persistence length $\ell_p$   to the length of a monomer,~\cite{bb} and $f$ is the same dimensionless measure of the force we have been using throughout.

Equation~(\ref{eq4.1})  is actually of the same form as our basic Hamiltonian, Eq.~(\ref{eq3.9}), but with an intra-chain Hamiltonian  ${\cal H}_0$  limited to nearest neighbor interactions.  To make this connection more literal, note that in a hard-sphere chain in any dimension, the angle $\theta$ between the bond linking a neighboring pair of spheres ${\hat \Omega}_{j-1}$  and the next bond ${\hat \Omega}_j$   is limited to  $0 <  \theta < \frac{2\pi}{3}$ by the excluded-volume interaction of the spheres.  In other words, the physical repulsion between any three consecutive spheres is equivalent to an interaction between the neighboring ``spins” that disfavors anti-parallel arrangements
\be
-\frac{1}{2} \le {\hat \Omega}_{j-1} \cdot {\hat \Omega}_j,
\ee
in much the same way that each ferromagnetic $ {\hat \Omega}_{j-1} \cdot {\hat \Omega}_j$  interaction term does in Eq.~(\ref{eq4.1}).  In fact, if one looks only at the interactions involving two neighboring bonds, one finds that the two models are quantitatively similar when $K$ is on the order of 1.  In two dimensions, for example, neighboring spin correlations are always ferromagnetic
\be
\langle {\hat \Omega}_{j-1} \cdot {\hat \Omega}_j \rangle_{\rm hard-sphere\,\,chain} = \frac{3}{2\pi} \int_0^{ \frac{2\pi}{3}} d\theta \cos{\theta} > 0
\ee
\be
\langle {\hat \Omega}_{j-1} \cdot {\hat \Omega}_j \rangle_{\rm worm-like\,\,chain} =  \frac{\int_0^{\pi} d\theta \cos{\theta} e^{K \cos{\theta}}}{ \int_0^{\pi} d\theta \,e^{K \cos{\theta}} } > 0
\ee
and are actually equal when $K = 0.90$. 

 Nonetheless the differences created by long-ranged interactions are profound.  The discretized worm-like chain is equivalent to a classical $n$-vector spin chain, a problem for which one can work out a number of key features analytically.~\cite{nn,oo,pp} It is well established, for example, that the susceptibility, Eq.~(\ref{eq2.10}), for such a chain tends to a constant in the infinite chain limit
 \begin{eqnarray}
 \chi \xrightarrow{N \to \infty} \frac{1+u}{1-u} = O(N^0) \label{eq4.2}\\
 u = I_1(K)/I_0(K) \,\,\,\ (2d)  \nonumber \\
 u = \coth(K) - \frac{1}{K} \,\,\,\,  (3d) \nonumber
 \end{eqnarray}
meaning that, despite the similarity in local interactions with the hard-sphere chain, worm-like chains have ideal polymer conformational statistics in all dimensions. Their root-mean-square end-to-end distances are always those of conventional random walks
\be
\langle {\bf R}^2 \rangle^{1/2} = \sqrt{Nd^2 \chi } = O(N^{1/2}) .
\ee

Given this behavior of the susceptibility, our discussion in the last section implies that large-$N$ worm-like chains should not only behave in a generally Gaussian fashion, the leading-order elasticity should be linear, even in two dimensions.  Equation~(\ref{eq3.17}) predicts specifically that the non-Gaussian parameter should be negative and vanish as $1/N$,
\be
\alpha_{NG} \sim - \frac{\chi^2}{N} = O(N^{-1} ).
\ee
 Indeed, this behavior is exactly what one sees in this model.  In the two-dimensional case, one can use transfer matrix ideas to compute the exact $2^{nd}$ and $4^{th}$ order susceptibilities needed for Eq.~(\ref{eq3.13}) and show (after considerable algebra)~\cite{qq} that
\begin{eqnarray}
\alpha_{NG}=  \frac{-2\chi + \chi'}{N} + O(N^{-2} )  \label{eq4.3}\\
 \chi' = \frac{1+u'}{1-u'},   \,\,\,\, u' = \frac{I_2 (K)}{I_0(K)} . \nonumber
 \end{eqnarray}
 
We are unaware of any way to do an analytical calculation of the full equation of state for the discretized worm-like chain (other than in one dimension, where the model is identical to the 1d Ising model). We shall therefore be content with comparing our predictions for the equation of state with simulations in the next section.  However, before we do so, there is an additional feature of the worm-like chain equation of state we should mention.  Worm-like chains have the intriguing property of exhibiting two different kinds of asymptotic behavior under the influence of large applied forces.  With weak elastic forces, the equation of state reverts to Eq.~(\ref{eq2.8}), the freely-jointed formula, under large tension, as it must, but when the chain becomes stiff enough, the behavior is qualitatively different.~\cite{cc}
	
	The standard way to view this distinction from the continuum perspective looks at the geometric mean of the persistence length $\ell_p$ and the Pincus length $\ell_{Pincus}$ (the thermal length scale defined by the magnitude of the applied force)~\cite{A,rr}
\be
\ell_{Pincus} = \frac{k_BT}{\rm applied\,\, force} = \frac{d}{f}. \label{eq4.4} 
\ee
In two dimensions, for example, when the geometric mean of the Pincus and persistence lengths is much less than the link size $d$, the asymptotic deviation from full extension is inversely proportional to the applied force
\be
(1-x) \to \frac{1}{2} \biggr( \frac{\ell_{Pincus}}{d} \biggr), \label{eq4.5} 
\ee
identical to what we wrote in Eq.~(\ref{eq2.8}).  But when $ \sqrt{\ell_{Pincus} \,\, \ell_p}  \gg d$ , that same deviation shows up as being inversely proportional to the square root of the force
\be
(1-x) \to \frac{1}{4} \sqrt{\ell_{Pincus} /\ell_p} \,\,\, . \, \label{eq4.6} 
\ee

This same dichotomy shows up in the discretized worm-like chain.  One can prove that when $N$ is large, the exact two-dimensional asymptotic equation of state associated with the Hamiltonian of Eq.~(\ref{eq4.1}) is~\cite{ff}
\be
(1-x) \to \frac{1}{2} \frac{1}{\sqrt{f^2 + 4fK} },  \label{eq4.7} 
\ee
which reverts to Eqs.~(\ref{eq4.5}) and (\ref{eq4.6}) when $f \gg K$ and $f \ll K$, respectively.

Our principal concern, of course, is not worm-like chain polymers, but polymers that have long-ranged excluded-volume forces.  
So, why are we bothering to discuss the different possible worm-like chain asymptotic behaviors?  The reason is that our entire approach to thinking about the effects of these long-ranged forces relied on incorporating the asymptotic behavior of our polymers in the absence of intra-chain forces.  However, one could imagine that a more accurate treatment might also build in the way in which asymptotic behavior is modified by the presence of short-ranged forces – which is precisely the information that Eq.~(\ref{eq4.7}) provides.
So at least for two-dimensional polymers, suppose we were to say that our limiting behavior at full extension is not the freely-jointed-chain-based formula shown in 
Eq.~(\ref{eq2.13}) (specialized to $D = 2$), but rather the worm-like chain generalization derived by expanding Eq.~(\ref{eq4.7})
\be
\big( 1-x(f) \big) = \frac{1}{2f} - \frac{K}{f^2} + \ldots
\ee
and then inverting to get  $f(x)$
\be
f(x) = \begin{cases} \bigr(\frac{2}{\chi}\bigr)x &x \to 0 \\ \frac{1}{2} \frac{1}{1-x} + \gamma + \epsilon(1-x)  &x \to 1\,\,\, . \end{cases} \label{eq4.8}
\ee

Both of the constants $\gamma$ and $\epsilon$ depend on the local interaction strength $K$ in known ways, but much as we did with the constant $\eta$  in Eq.~(\ref{eq2.13}), we can imagine treating these two constants as parameters whose values we expect to be modified by the long-ranged interactions, and therefore as values to be determined from simulation.  The simplest Pad\'e form for the equation of state that is consistent with both Eq.~(\ref{eq4.8}) and the necessary ${\bf R}  \to -{\bf R}$ symmetry is then a straightforward generalization of Eq.~(\ref{eq2.14})
\begin{eqnarray}
f(x) &=& \biggl(\frac{2}{\chi}\biggr) \frac{x + Ax^3 + Bx^5 + Cx^7}{1-x^2} \label{eq4.9} \\
A &=& -3 + \biggl(\frac{\chi}{2}\biggr) \biggl (\frac{59}{16} + \frac{5}{2} \gamma +\frac{1}{2} \epsilon \biggr) \nonumber \\
  B &=& 3+  \biggl(\frac{\chi}{2}\biggr) \biggl (-\frac{66}{16} -4\gamma -\epsilon \biggr) \nonumber \\
  C &=& -1+  \biggl(\frac{\chi}{2}\biggr) \biggl (\frac{23}{16} + \frac{3}{2} \gamma +\frac{1}{2} \epsilon \biggr). \nonumber  
\end{eqnarray}
This revision turns out not to change our scaling predictions for the non-Gaussian parameter, Eq.~(\ref{eq3.17}), because the leading-order nonlinearity we predict is still given by Eq.~(\ref{eq3.16}), and because we still expect  $A \sim \chi$  for large $N$.  The equation of state itself, though, could have significant quantitative differences.  We will examine this possibility in the next section.

As a final, parenthetical note, we should mention that when very stiff polymers
are modeled with worm-like chains, the persistence length can be of the order of the
polymer’s contour length, which makes $K$ itself of $O(N)$
and invalidates most of the
predictions for the  $N$ dependence discussed in this section. In fact, our large-deviation theory is
inapplicable in this regime because the constant-force and constant-extension equations
of state are noticeably distinct.~\cite{t, ss, tt, uu} However, polymers that
are this stiff are completely insensitive to the long-ranged excluded-volume effects that
the theory is designed to treat.

\section{COMPARISON OF SIMULATED POLYMER BEHAVIOR WITH OUR LARGE-DEVIATION PREDICTIONS}\label{sV}

\subsection{Simulation methods} 

The principal simulations we need to carry out are those for the hard-sphere chains under an applied tension, the system described at the beginning of Sec.~\ref{sIV}.  We will examine isolated two- and three-dimensional chains over a range of $N$ values, but our focus will be on chains with $N = 1000$ links.

The hard-sphere calculations are performed using a Monte-Carlo (MC) simulation with reptation (sometimes called “slithering-snake” dynamics).~\cite{vv} The algorithm consists of choosing at random a sphere from the head or tail of the polymer and attempting to place it at the other end.  A spherically-symmetric trial location a distance $d$ from that far end is then computed.  If a sphere at this position would overlap with any other sphere in the polymer, the move is rejected.  Otherwise, the move is accepted provided it obeys the Metropolis criterion that $e^{f \Delta X} > \mu$.  Here $\mu$ is a uniformly chosen random number in $[0,1)$, $f$ is the magnitude of the applied tensile force, and $\Delta X$ is the change in the end-to-end displacement length of the polymer in the direction of the applied force.  For $N+1$-sphere chains, we define a “step” in the algorithm to be $N+1$ such attempted moves.

We took our initial configuration to be a straight polymer.  A more random initial configuration would be problematic in two dimensions because the head or tail of the polymer can find itself in a cul-de-sac, potentially preventing the polymer from equilibrating.  For large values of $f$, we performed 100,000 MC steps for equilibration and then averaged over 200,000 MC steps.  For small values of $f$, where the error bars are significant, we used 50 independent runs of 100,000 MC equilibration steps and 100,000 MC steps for averaging so we could provide error bars showing the standard error of the mean of the 50 runs.

While the hard-sphere model contains all of the features of actual polymers that we wanted to study in this paper, as we mentioned in the last section, we also want to compare with the behavior of worm-like chains and freely-jointed chains.  The two-dimensional discretized worm-like chain model we use is equivalent to the one-dimensional XY model, Eq.~(\ref{eq4.1}), with $K=1.00$.  That model has no non-nearest-neighbor interactions, so it is simulated via standard Metropolis MC methods using periodic boundary conditions~\cite{vv1} and $N$ = 1000 spins.  The properties of the freely-jointed chain are calculated using the exact expressions given in Eq.~(\ref{eq2.7}).
	
	As we emphasized throughout our discussion, the formalism that we derived in this paper, and that we would like to test with these simulations, actually requires some (limited) information from these same simulations. The necessary susceptibilities $\chi$ were computed just by using the definition, Eq.~(\ref{eq2.10}).  With the hard-sphere simulations, we found $\chi = 1.09 N^{0.49}$ in 2D and $\chi = 1.45N^{0.20}$ in 3D for $200 \leq N \leq 1600$, consistent with the Flory result $\chi \sim N^{2\nu-1}$ with $\nu = 3/(D+2)$ (which is exact in 2D and close to the exact result $\nu \approx 0.588$ in 3D).~\cite{h}

Finding the parameters $\eta$, $\gamma$, and $\epsilon$, though, required fits to the asymptotic region of the simulated equations of state.  To compute $\eta$ for the hard-sphere chain, we used a least-square fit of $f$ versus $1 \over (1-x)$, as in Eq. (\ref{eq2.13}) for $x \to 1$, with values of $f  > 10$.  The fitted slope was within $1\%$ of $(D-1)/2$, as desired.  We found $\eta = 0.196$ in 2D and $-0.221$ in 3D for $N = 1000$ using four-point fits, but the specific values should not be taken too literally because they vary greatly depending on how many points are used in the fits.  Examination of $\eta$ for higher $N$ simulations also led to similarly variable outcomes, but with magnitudes in all cases less than 1 – consistent with the assumptions made in our scaling discussion in Sec. III.

To compute the parameters $\gamma$ and $\epsilon$ for the 2D worm-like chain generalization given in Eq.~(\ref{eq4.9}), we minimized the difference between Eq.~(\ref{eq4.9}) and the simulated hard-disk data, giving us $\gamma = 0.97$ and $\epsilon = -10.11$

\subsection{Elastic and inelastic behavior of polymers}

\begin{figure}
    \includegraphics[width=0.8\linewidth]{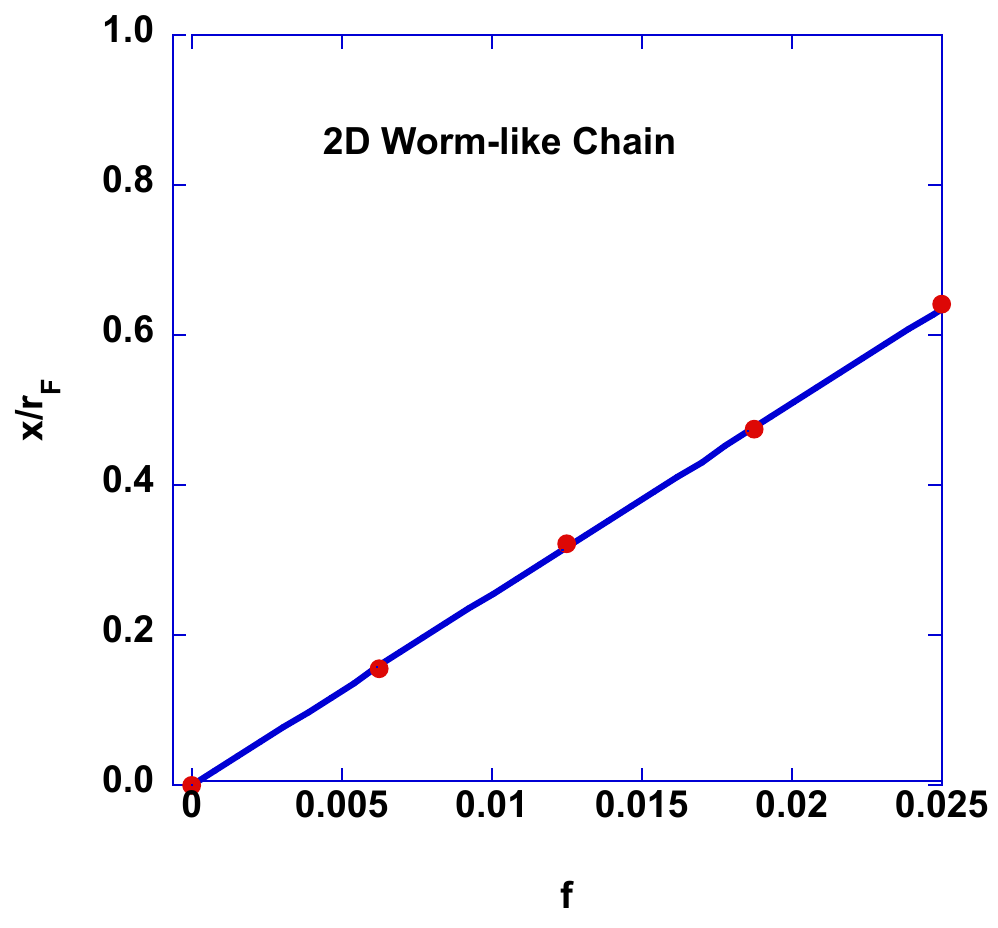}
    \caption{Equation of state of the discretized worm-like chain model in two dimensions (the 1d classical XY model) at small values of the applied force $f$ and with a monomer-sized persistence length ($K = 1.00$). The dots show the results of Monte Carlo simulations with $N = 1000$, and the line represents our large-deviation theory prediction, Eq.~(\ref{eq2.14}).  In two dimensions, and at small $f$ values, Eqs.~(\ref{eq5.1}) and (\ref{eq2.13}) imply $x/r_F = {\sqrt{N \chi} \over 2} f$. From the simulation for $N=1000$, $\chi (f=0) = 2.60$, so that theory predicts $x/r_F = 25.5 f$, which is very close to the fitted slope of the data ($25.7$).  (Analytical predictions from Eq.~(\ref{eq4.2}) also match these values, yielding $\chi(f = 0) = 2.613$ and a corresponding slope of 25.56.) Data was collected from 10 independent runs, each averaged over 50,000 MC steps after 5,000 steps of equilibration.  The error bars are smaller than the symbol size on this plot. }
        \label{f1}
\end{figure}

\begin{figure}
    \includegraphics[width=0.8\linewidth]{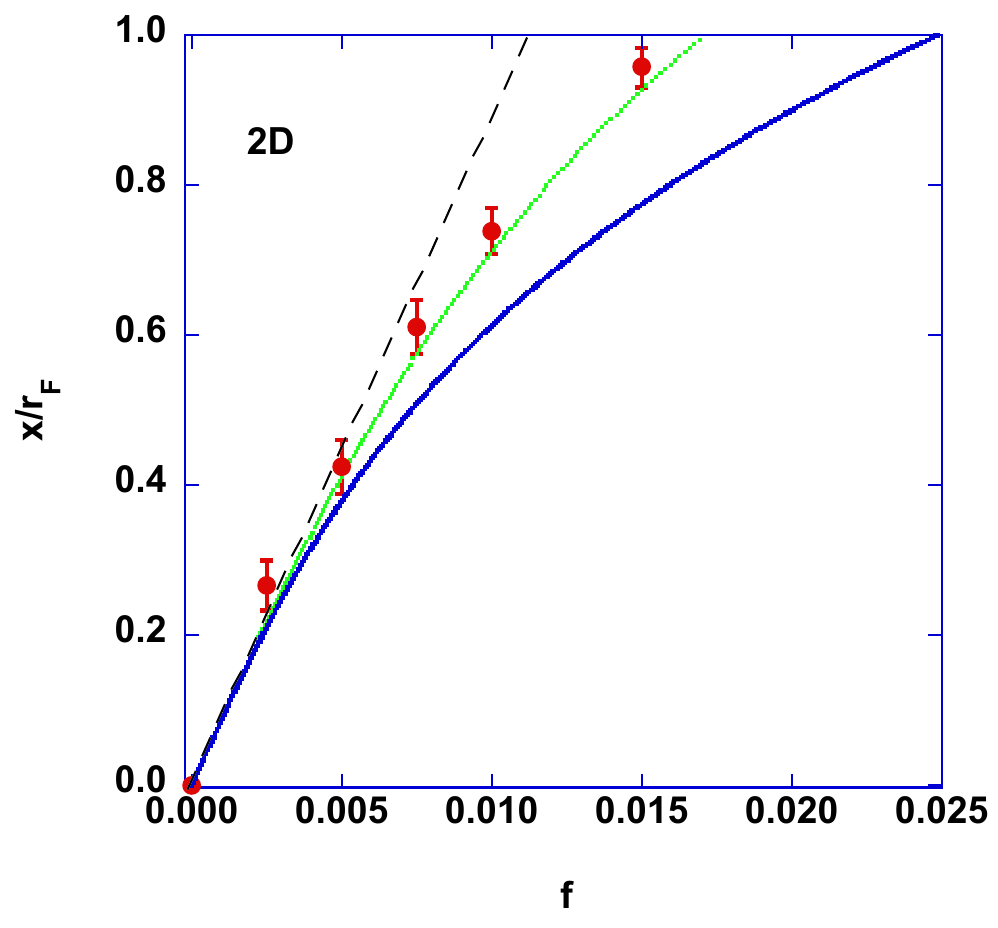}
    \caption{Equation of state for hard-disk polymers in two dimensions at small values of the applied force $f$.  Monte Carlo hard-disk-polymer data (dots) with $N = 1000$ are compared with the linear elastic prediction based on the small $f$ data (dashed line), the 5th order large-deviation theory (Eq.~\ref{eq2.14}) (lower solid-blue curve), and the 7th order large-deviation theory (Eq.~\ref{eq4.9}) (upper solid-green curve). }
    \label{f2}
\end{figure}

\begin{figure}
    \includegraphics[width=0.8\linewidth]{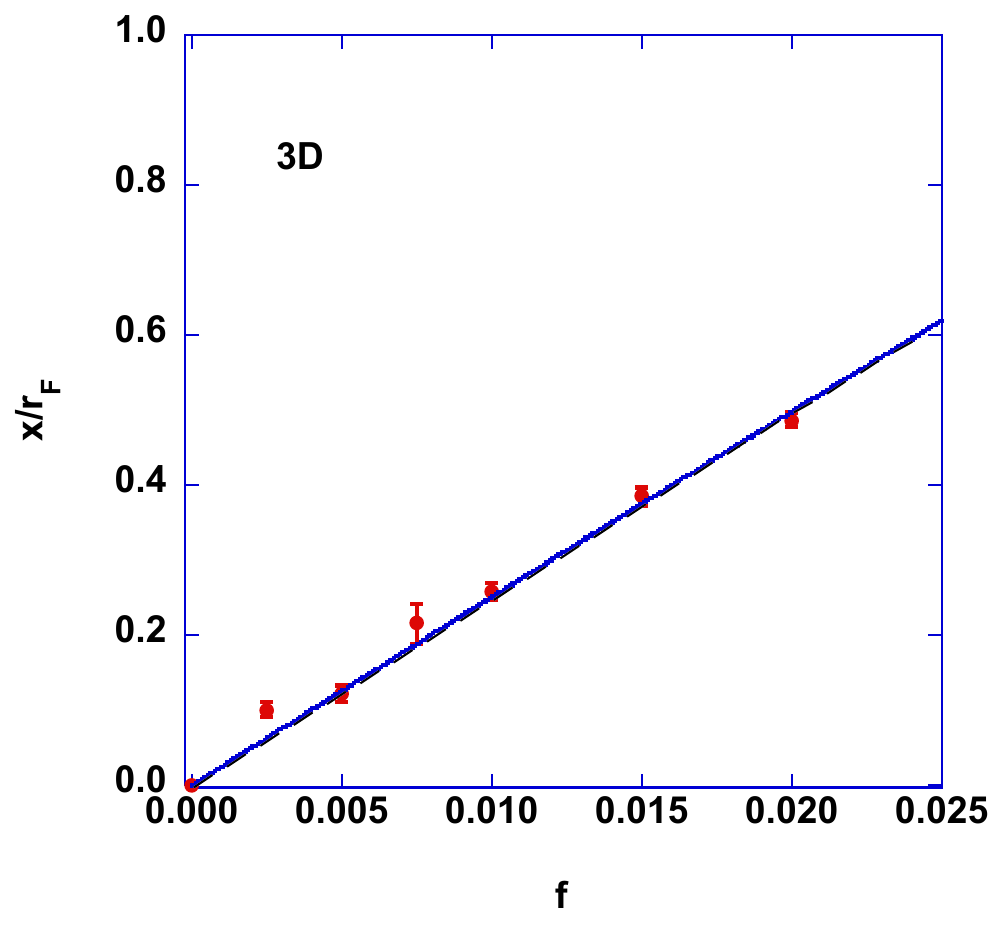}
    \caption{Equation of state for hard-sphere polymers in three dimensions at small values of the applied force $f$.  Monte Carlo hard-sphere polymer data with $N = 1000$ (dots) are compared with the linear elastic prediction based on the small $f$ data (dashed line), and both the $5^{th}$ and $7^{th}$ order large deviation theories (solid lines).  All three theoretical curves are essentially indistinguishable at this scale.}
    \label{f3}
\end{figure}

We begin by looking at the conformations of polymer chains in the putatively elastic regime, those with end-to-end distances close to where they would be in the absence of any applied forces.  What we plot versus the scaled applied force $f$  (Eq.~(\ref{eq2.4})) in Figs.~\ref{f1}-- \ref{f3} is the ratio of $x$, the scaled average projection of the polymer’s end-to-end vector $\bf R$ in the direction of that applied force, to the scaled Flory distance $r_F$.  Equations~(\ref{eq1.1}), (\ref{eq2.4}), and (\ref{eq3.1}) show this ratio to be
\be
\frac{x}{r_F} = \frac{\langle {\bf R} \cdot {\hat f} \rangle_f }{ \langle {\bf R}^2 \rangle^{1/2} } = \sqrt{\frac{N}{\chi}} x\, . \label{eq5.1}
\ee

	Polymers that are genuinely elastic should generate linear plots on this scale; they should obey Hooke’s law.  Indeed, what we see for an $N = 1000$ two-dimensional worm-like chain, Fig.~\ref{f1}, and an $N = 1000$ three-dimensional hard-sphere chain, Fig.~\ref{f3}, is precisely the linear, elastic behavior predicted in Secs. III and IV.  But, as we also predicted in Sec.~\ref{sIII}, an $N = 1000$ two-dimensional hard-disk chain, Fig.~\ref{f2}, exhibits a curvature indicative of a clear nonlinear elasticity when viewed on the same scale.  The dashed lines in Figs.~\ref{f2} and \ref{f3}, which we show for comparison, are simulated linear elastic result derived from a straight-line fit to the smallest $f$ points in the simulated data. The solid curves are predictions based on the large-deviation-theory and will be discussed shortly.

\begin{figure}
    \includegraphics[width=0.8\linewidth]{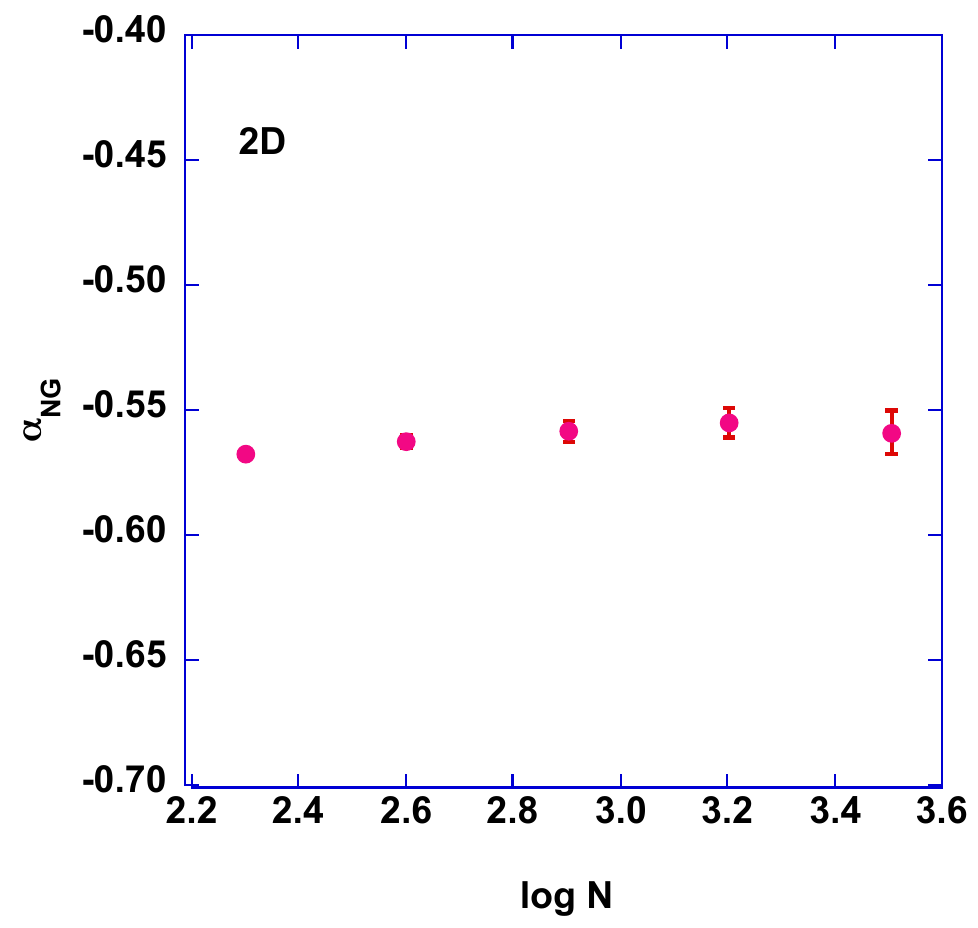}
    \caption{Non-gaussian parameter $\alpha_{NG}$, Eq.~(\ref{eq3.7}), computed from Monte Carlo simulations of two-dimensional hard-disk polymers over a range of $N$ values.  Each data point is from 50 independent runs averaged over 200,000 MC steps after 50,000 equilibration steps.}
    \label{f4}
\end{figure}

	A more quantitative implication of the scaling analysis in Sec.~\ref{sIII}  is that two-dimensional polymers with long-ranged excluded volume should have their non-Gaussian parameter $\alpha_{NG}$   be a finite negative number independent of $N$, (Eqs.~(\ref{eq3.17}) and (\ref{eq3.18})).  We test this supposition in Fig.~\ref{f4} by explicitly computing  $\alpha_{NG}$  from our two-dimensional hard-sphere chain simulations using Eq.~(\ref{eq3.7}) over a range of $N$ values
( $200 \le N \le 3600$) and showing that this expectation is indeed verified to within our numerical precision.  Evaluating  $\alpha_{NG}$  when the elasticity is closer to linear is more challenging because of the difficulty in sampling something as noisy as a $4^{th}$  order susceptibility, but we have also verified that the simulated value of  $\alpha_{NG}$  for a two-dimensional worm-like chain with $N = 1000$ and $K = 1$ is in at least rough agreement with the much smaller value of $-3.99 \times 10^{-3}$  predicted by 
Eq ~(\ref{eq4.3}).  The same analysis further predicts that in three dimensions, the hard-sphere chain  $\alpha_{NG}$ should also vanish with increasing $N$, but quite slowly.  Our three-dimensional hard-sphere chain simulations were unable to see any such decay with $N$, but the linearity evident in Fig.~\ref{f3} seems to support the basic premise that three-dimensional polymers have substantial Gaussian character by the time $N = 1000$.

\subsection{Equations of state of polymers} 

The behavior of the polymer equation of state beyond the small-extension regime has its own interest,~\cite{a,b,ww} so we have looked at a variety of polymer systems and compared the equations of state with our large-deviation prediction, Eq.~(\ref{eq2.14}).
	
\begin{figure}
    \includegraphics[width=0.8\linewidth]{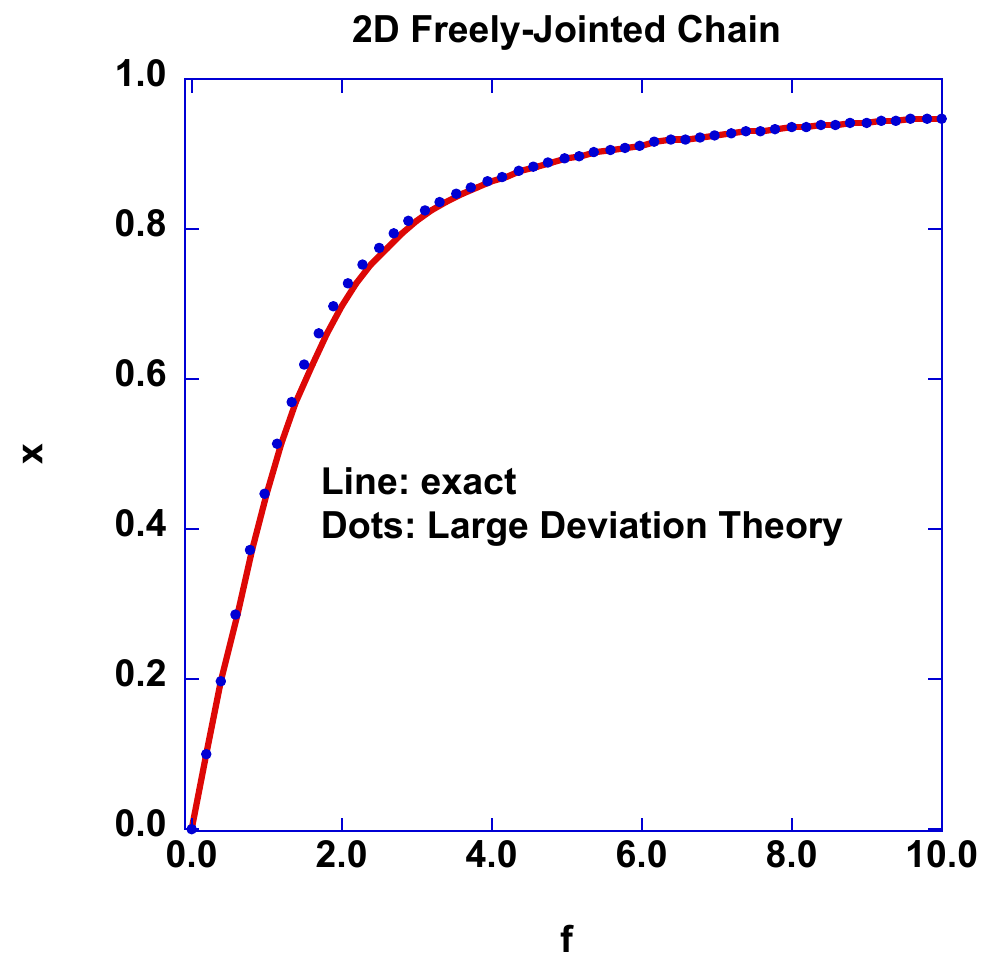}
    \caption{Equation of state for freely-jointed polymers in two dimensions.  The solid line gives the exact result (Eq.~\ref{eq2.7}) for the 2d freely-jointed chain, and the dots present the large-deviation theory predictions, Eq.~(\ref{eq2.14}), with $\chi = 1$ and $\eta = (3-D)/4= 1/4$, equal to their ideal values.}
    \label{f5}
\end{figure}

\begin{figure}
    \includegraphics[width=0.8\linewidth]{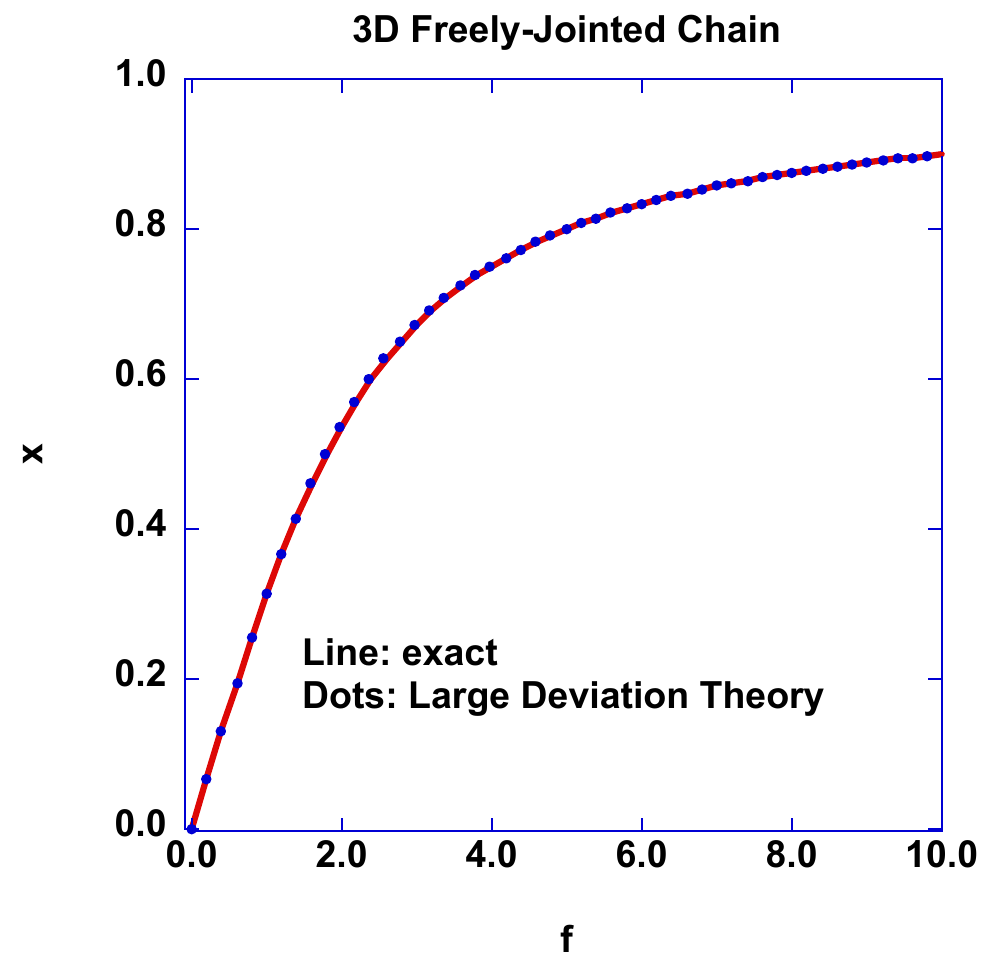}
    \caption{Equation of state for freely-jointed polymers in three dimensions.  The solid line gives the exact result (Eq.~\ref{eq2.7}) for the 3d freely-jointed chain, and the dots present the large-deviation theory predictions, Eq.~(\ref{eq2.14}), with $\chi = 1$ and $\eta = (3-D)/4= 0$ equal to their ideal values.}
    \label{f6}
\end{figure}

	The equation of state when the polymer links are all independent (the freely-jointed chain) is analytically known, Eqs.~(\ref{eq2.6}) and (\ref{eq2.7}), but it provides a useful baseline check on the development in this paper.  In the freely-jointed case, one knows from Eq.~(\ref{eq2.10}) that $\chi = 1$ and from asymptotic expansion of the Bessel functions~\cite{u} in Eq.~(\ref{eq2.6}) that $\eta$, the  $O(N^0)$  term in Eq.~(\ref{eq2.8}) is $(3-D)/4$.  As shown in Figs.~\ref{f5} and \ref{f6}, using these values in our large-deviation form for the equation of state provides a quantitatively accurate representation of the full range of behavior under applied tension (from elastic to saturation) in both 2 and 3 dimensions.
	
\begin{figure}
    \includegraphics[width=0.8\linewidth]{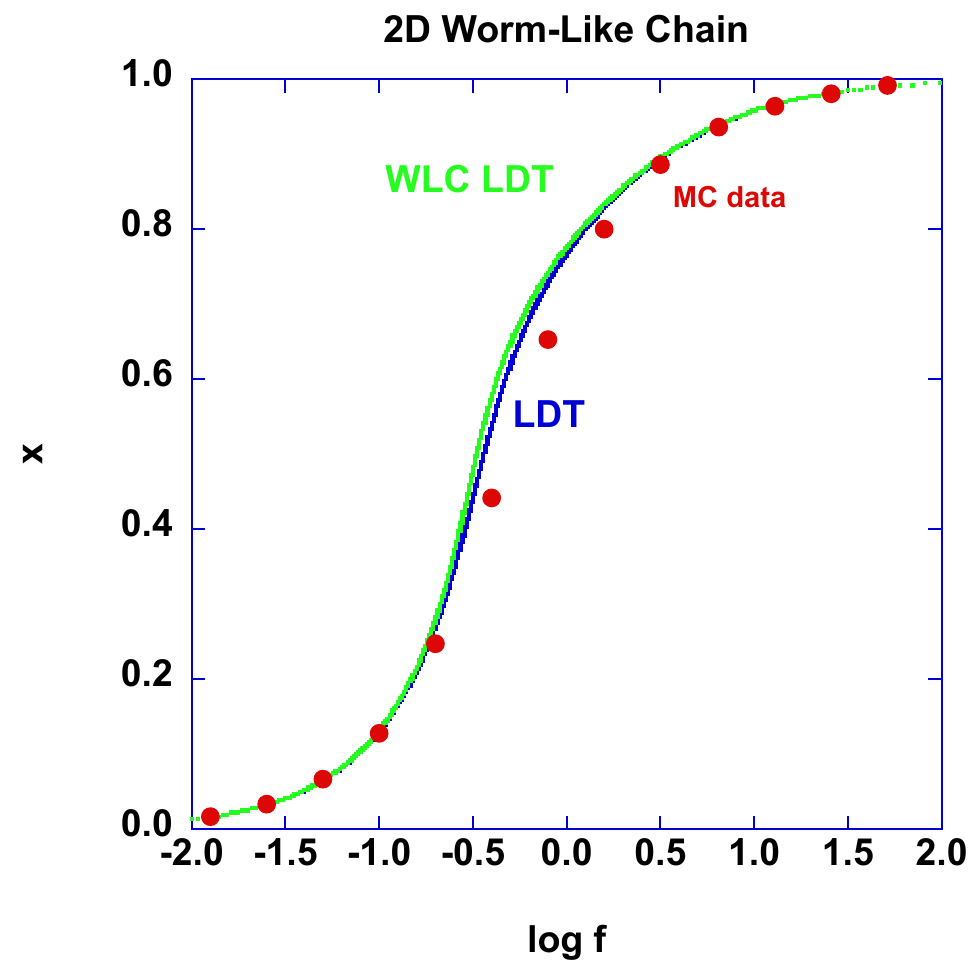}
    \caption{Equation of state of the discretized worm-like chain model in two dimensions (the 1d classical XY model) with a monomer-sized persistence length ($K = 1.00$).  The dots present the results of Monte Carlo simulation with $N = 1000$. The curves show the predictions of the 5th order large-deviation theory, (Eq.~\ref{eq2.14}), (lower blue curve) and the 7th order large-deviation theory, (Eq.~\ref{eq4.9}) (upper green curve).  Error bars are smaller than the symbol size on this plot.}
    \label{f7}
\end{figure}

	The situation in which the polymer links have local (but not long-ranged) interactions, the discretized worm-like chain, Fig.~\ref{f7}, also seems to reveal reasonably satisfactory agreement between our original large-deviation formula, Eq.~(\ref{eq2.14}) and Monte Carlo simulated equation of state.  In this and subsequent figures, we have plotted the scaled projected end-to-end distance versus the logarithm of the applied force in order to highlight whatever errors our formalism might make in the transition regime between elastic and saturation behavior.  Despite the presence of some deviations, the predictions seem quite robust; they show no appreciable change when we switch to the (presumably improved) large-deviation formula we constructed specifically with the worm-like chain in mind, Eq.~(\ref{eq4.9}).
	
\begin{figure}
    \includegraphics[width=0.8\linewidth]{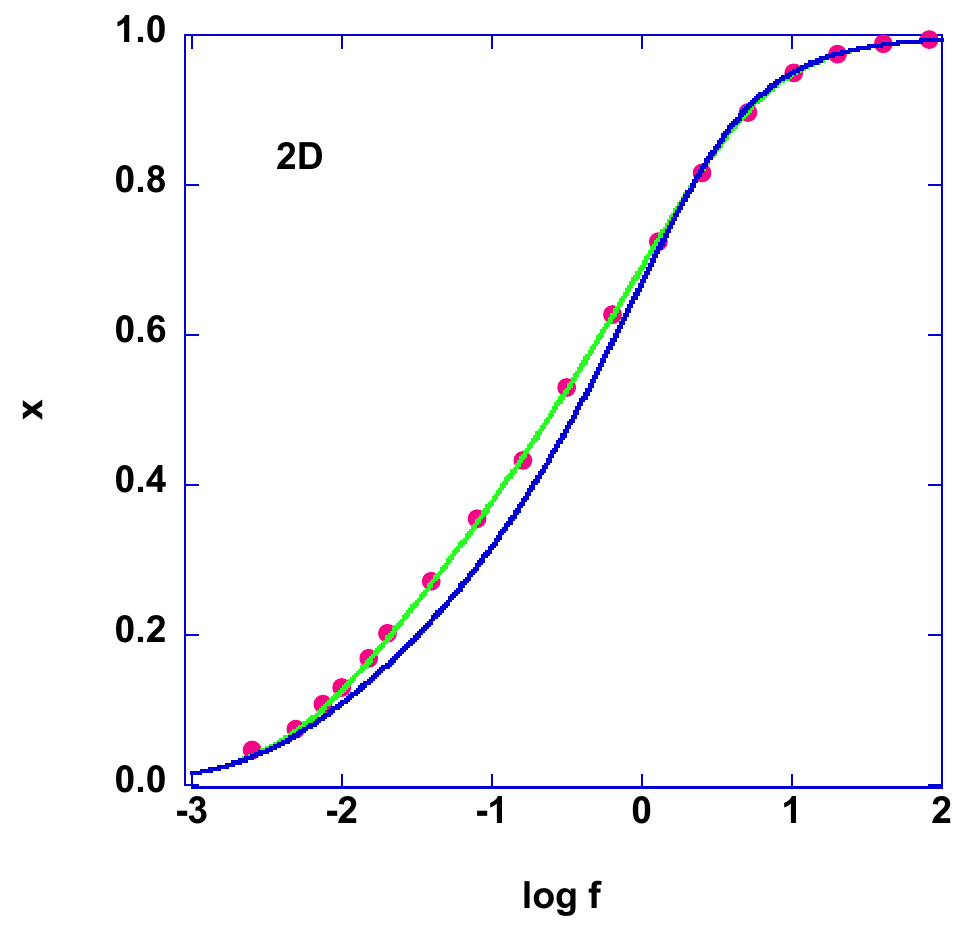}
    \caption{Equation of state of hard-disk polymers in two dimensions with $N = 1000$ monomers.  The dots present the results of Monte Carlo simulation. The curves show predictions of the $5^{th}$ order large-deviation theory, (Eq.~\ref{eq2.14}), (lower blue curve) and the $7^{th}$  order large-deviation theory, (Eq.~\ref{eq4.9}) (upper green curve).  Error bars are smaller than the symbol size on this plot.}
    \label{f8}
\end{figure}

\begin{figure}
    \includegraphics[width=0.8\linewidth]{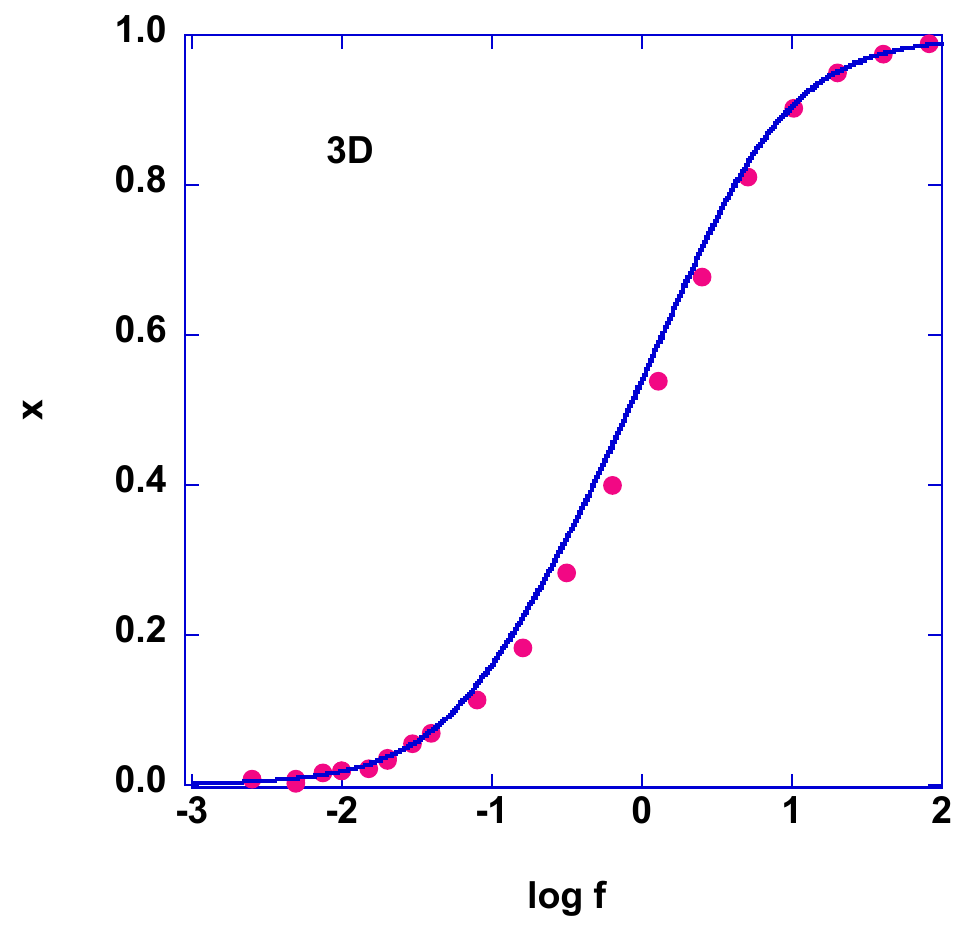}
    \caption{Equation of state of hard-sphere polymers in three dimensions with $N = 1000$ monomers.  The dots present the results of Monte Carlo simulation. The blue curve shows predictions of the $5^{th}$ order large-deviation theory, (Eq.~\ref{eq2.14}).  Error bars are smaller than the symbol size on this plot.}
    \label{f9}
\end{figure}

	Finally, consider the situation we are really targeting in this paper: polymers with realistic long-ranged excluded volume interactions.  We compare the simulated equations of state of hard-sphere chain in two and three dimensions with the results of our basic large-deviation formula, Eq.~(\ref{eq2.14}), in Figs.~\ref{f8} and \ref{f9}.  The basic large-deviation formula seems to perform quite respectably in three dimensions, both when we examine it over the full range of behaviors in Fig.~\ref{f9} and when we return to look at the small-extension regime, Fig.~\ref{f3}.  However, in two dimensions, we find that the basic large-deviation formula (blue) curves, while qualitatively correct, are noticeably distinct from the simulated results.  Returning to Fig.~\ref{f2}, in particular, shows that this formula clearly overestimates the small-extension nonlinearity.  Interestingly, though, Eq.~(\ref{eq4.9}), the improved large deviation formula shown in the green curves, produces what is apparently quantitative agreement with the simulation at all length scales, as shown in Figs.~\ref{f2} and \ref{f8}.
	
	It is unclear precisely why the more accurate inclusion of local forces in Eq.~(\ref{eq4.9}) should become more relevant when long-ranged forces come into play, but it is possible that the precise level of small-extension nonlinearity that these forces generate is exceptionally sensitive to local interactions in two dimensions.  Evidently, local and long-ranged forces act in a somewhat unexpectedly cooperative fashion when a polymer is confined to two dimensions.

\section{Concluding Remarks}\label{sVI}

The main point of this work was to show that the analytically available information on polymer conformational statistics near the largest possible extensions is a surprisingly valuable aid in deducing free energies from polymer simulations, even when the interest is in behavior far from that regime.  By combining relatively-easy-to-access simulation results on the deviations from ideal behavior in the large-extension regime (that generated by strong applied forces) with simulated data from the opposite, highly non-ideal, small-extension (weak-applied-force) domain, we were able to construct accurate interpolated force-extension relations.  Large deviation theory then gave us a simple route to turn those simulation-derived relations into statements about how the free energy depends on polymer extension.  For example, in our hard-sphere chain examples what we were really calculating, with little effort, was the decrease in entropy of self-avoiding random walks at larger and larger net walk displacements – what in the comparable lattice problems would be a nontrivial calculation of how the number of self-avoiding walks varies with net displacement.~\cite{A,xx}

	The particular feature we highlighted in this paper, the amplified role of self-avoidance in two-dimensional polymers, was largely a demonstration project. But, we should point out that this sample problem is not totally without experimental relevance.~\cite{yy,zz,aaa}  For one thing, polymer force-extension relations (what we have called the polymer ``equations of state”) are now experimentally accessible in a number of ways, including measurements as a function of hydrodynamic forces (solvent flow), electric fields, and various ways of microscopically pulling on the ends of the polymer.~\cite{o,r,bbb}  Truly two-dimensional situations are harder to come by.~\cite{d,ccc}  But polymers restricted to topologically two-dimensional regions, such as those confined to lie within lipid membranes or in narrow slits are, in fact, experimentally accessible.~\cite{yy,zz,bbb}

In any case, our formalism did lead us to some new insights into the onset of nonlinear elasticity in (literally) two-dimensional polymers. Our approach made it easy to see how the relative finite-size scaling of the quadratic and quartic terms in the free energy depends on whether there are long-ranged excluded volume interactions and on what the spatial dimensionality is – and that it was only when we have both long-ranged excluded volume effects and polymers confined to two dimensions that the nonlinear quartic contributions can survive the large-chain limit.  We were able to confirm this finding numerically by comparing hard-disk polymers with discretized worm-like chains, allowing us to discriminate between long-and short-ranged excluded volume effects, and by comparing hard-disk and hard-sphere chains, letting us pinpoint the effects of dimensionality.

However, the overall intent in this paper was to present a fairly general approach to computing the free energy dependence on polymer extension from simulations, one that could readily be applied to more complex polymer structures and environments than we have looked at here, including dilute solutions, melts, polymers with collapse transitions, copolymers, and heteropolymers such as peptides and nucleic acids.~\cite{a} The basic portrait of polymers as being built out of discrete, coupled orientational degrees of freedom, in tandem with our large-deviation framework, allows one to think, in particular, about finite-size effects in microscopically-detailed liquid-state polymeric systems.  That should allow us to explore phenomena sensitive to intra-chain heterogeneity, for example.~\cite{ddd}

Having made these points, we should also acknowledge that this method has some notable limitations.  It does not seem to be well suited for such routine non-order-parameter-like quantities as the radius of gyration and the structure factor, for example.  For us, the term “order parameter” means a sum over contributions from individual microscopic degrees of freedom, with the important proviso that the large-order-parameter limit makes those degrees of freedom become independent.  The structure factor and radius of gyration depend on sums over individual polymer sites (rather than bonds), so the chain connectedness generates a strong degree-of-freedom/degree-of-freedom correlation between sites that is always present.  In addition, our insistence on restricting permissible polymer models to those with just orientational degrees freedom (so that the polymer’s end-to-end length saturates at some maximum value) precludes using standard harmonic-bond descriptions~\cite{s} and actually introduces differences in probability densities from more faithful descriptions of intramolecular forces.~\cite{eee}
	
	Nonetheless, it is worth asking just how well our methodology does within its domain of validity.  Since our method interpolates between simulation-derived large-extension behavior and small-extension behaviors, the real test of the method is its ability to predict the polymer extension quantitatively in the intermediate regimes.
	
	To be precise, ``small-extension” and ``large-extension” in our context means that the end-to-end distance $R$ is either very small or very large compared to the polymer’s contour length $L$.  But, as we discussed in Sec.~\ref{sIII}, it is possible for $R$ to be small compared with $L$ without being small compared with the root-mean-square length $R_F$.  In particular, it was by looking at situations in which $R/R_F$ is $O(1)$ that told us about the unique nonlinear small-$R$ elasticity of two-dimensional polymers.  A more demanding test of our methods, though, is to go into the genuinely intermediate regime in which $R/L$ has more moderate values ($R/R_F > 1$, $R/L \ll 1$).  There it is well known that the free energy $F(R)$ scales as~\cite{A}
\be
F(R) = \biggl(\frac{R}{R_F}\biggr)^{\delta}\,\,\,\,\,  \delta = \frac{1}{1-\nu} \begin{cases} 4 \,\,\, \,\,\,\,\,D = 2\\ 2.5\,\,\, D = 3\\ 2 \,\,\,\,\,\, \,\,D = 4\,\,\, . \end{cases} 
\ee
	 		 
Our formalism does not explicitly predict this scaling (other than correctly predicting Gaussian behavior in the ideal ($D = 4$) case).   However, as one can see from the simulation/prediction agreement at intermediate $x$ values in Figs.~\ref{f8}  and \ref{f9}, our numerical results must be roughly consistent with the corresponding scaling relation for the equation of state\cite{fff}
\be
f(x) = x^{\delta-1}. 
\ee
	  
We look forward to seeing whether this level of agreement continues to hold for more realistic and more challenging polymer problems.

\begin{acknowledgments}
	We thank Brown University and the American Hellenic Educational Progressive Association for support provided to Eleftherios Mainas.
\end{acknowledgments}

\section*{AUTHOR DECLARATIONS}

\subsection*{Conflict of Interest}

		The authors have no conflicts to disclose.

\subsection*{Authors' Contributions}

Eleftherios Mainas: Conceptualization (equal); Methodology (equal).

Jan Tobochnik: Conceptualization (equal); Methodology (equal); Software (lead); Visualization (lead); Writing (Original draft, review, and editing) (equal).

Richard Stratt: Conceptualization (equal); Methodology (equal); Writing (Original draft, review, and editing) (equal).

\section*{DATA AVAILABILITY}

The data presented in the figures shown here are available from the authors on request.


\begin{thebibliography}{88}
\bibitem{n} In this paper we will be using the term ``equation of state” to refer only to the force-extension relationship and not to the relationship between pressure and density.
\bibitem{t}Assuming the existence of a unique, well-defined, force-extension equation of state, $f = f(x)$, is tantamount to assuming that there are no quantitative differences between constant-force and constant-extension ensembles.  Our large deviation theory is therefore only applicable to situations in which these ensemble differences are negligible.
\bibitem{v}  In writing our asymptotic expression for the deviation from full extension, we have neglected terms exponentially small in $f$.  Note that in 3D, all but one of these contributions to the deviation are exponentially small, and that in 1D, all of the contributions are exponentially small.
\bibitem{x} This expectation is predicated on the fact that $\chi = O(N^{2\nu-1}) \gg1$ and on the assumption that  $\eta = O(N^0)$.  We numerically confirm that this supposition is correct in the simulations described in Sec.~\ref{sV}.
\bibitem{fff} While the boundaries of this intermediate regime are not all that well defined, and the fit details are sensitive to the precise region fitted, log-log plots for our hard-sphere chains, do, in fact, show that both the Monte Carlo (MC) data and the large-deviation theories (LDT) are fairly consistent with the expected scaling in this regime.  The expected $\delta -1  = 3$  in 2D, and sample fittings of both the MC and the improved LDT results yield slopes of approximately 3.3.  The expected $\delta -1 = 1.5$ in 3D, and sample fittings of both the MC and LDT results yield slopes of approximately 1.5.
\end{thebibliography}
\end{document}